# Topological and AIM analyses beyond the Born-Oppenheimer paradigm: New opportunities

Mohammad Goli and Shant Shahbazian[*]

*Faculty of Chemistry, Shahid Beheshti University, G. C. , Evin, Tehran, Iran, 19839, P.O. Box 19395-4716.*

Tel/Fax: 98-21-22431661

E-mail:
(Shant Shahbazian) chemist_shant@yahoo.com

[*] Corresponding author

**Abstract**

The multi-component quantum theory of atoms in molecules (MC-QTAIM) analysis is done on methane, ethylene, acetylene and benzene as selected basic hydrocarbons. This is the first report on applying the MC-QTAIM analysis on polyatomic species. In order to perform the MC-QTAIM analysis, at first step the nuclear-electronic orbital method at Hartree-Fock level (NEO-HF) is used as a non-Born-Oppenheimer (nBO) ab initio computational procedure assuming both electrons and protons as quantum waves while carbon nuclei as point charges in these systems. The ab initio calculations proceed substituting all the protons of each species first with deuterons and then tritons. At the next step, the derived nBO wavefunctions are used for the "atoms in molecules" (AIM) analysis. The results of topological analysis and integration of atomic properties demonstrate that the MC-QTAIM is capable of deciphering the underlying AIM structure of all the considered species. Also, the results of the analysis for each isotopic composition are distinct and the fingerprint of the mass difference of hydrogen isotopes is clearly seen in both topological and AIM analyses. This isotopic distinction is quite unique in the MC-QTAIM and not recovered by the orthodox QTAIM that treats nuclei as clamped particles. The results of the analysis also demonstrate that each quantum nucleus that forms an atomic basin resides within its own basin. The confinement of quantum nuclei within a single basin is used to simplify the basic equations of the MC-QTAIM paving the way for future theoretical studies.

# 1. Introduction

We are living in the age of ab initio quantum chemistry; the obstacles of solving the basic quantum equations are gradually disappearing and the state-of-art sophisticated computational procedures lead to the dawn of reliable predictions. However, this is just one face of the coin; the "chemistry" hidden within the complicated wavefunctions or reduced density matrices, composed of hundreds, thousands or even millions of the Slater determinants, are hard to comprehend. The happy days of "paper and pencil" methods, developed in the age of (semi-)qualitative molecular orbital methods, appealing to the golden rule for making wavefunctions as simple as possible, are now gone. The age of ab initio quantum chemistry needs its own apparatus to "extract" chemistry from the complicated wavefunctions; in a broad perspective, a complicated ab initio wavefunction is just a "code" and the principles of quantum mechanics are recipes disclosing how to derive physical observables from this code. However, the principles of quantum mechanics are silent on cracking the code in order to derive "chemical observables", e.g. electronegativity, atoms in molecules, hardness, etc. [1]. Thus, one needs to go beyond the formal quantum mechanics "designing" *machines* that are capable of using ab initio wavefunctions/density matrices as *inputs* and generating chemical observables as *outputs* [1].

One of these machines is the Quantum Theory of Atoms in Molecules (QTAIM) that aims to derive the atoms in molecules and their properties from ab initio wavefunctions [2-4]. The original formulation of the QTAIM was just confined within the Born-Oppenheimer (BO) paradigm, treating electrons as quantum waves and nuclei as clamped point charges, using the electronic wavefunctions as inputs [2]. The recently developed multi-component QTAIM (MC-QTAIM) goes beyond this paradigm and is capable of using non-adiabatic wavefunctions as inputs; treating nuclei as quantum waves instead of clamped nuclei [5-10]. Even more, this new formalism is capable of using the wavefunctions of the positronic and the muonic species as inputs yielding the atoms in molecules in these exotic species [11-14]. Since the MC-QTAIM reproduces the results of the orthodox QTAIM by tending the masses of quantum nuclei to infinity [6,7], practically the clamped nuclei limit, it is safe to claim that the former encompasses the latter. Accordingly, the MC-QTAIM is a unified scheme that may act as a machine to derive the atoms in molecules and their properties from large *types* of ab initio wavefunctions (BO based electronic wavefunctions, non-Born-Oppenheimer (nBO) wavefunctions including multiple quantum and clamped particles, …).

While in our previous computational studies only diatomic species were considered [5-14], in present contribution the first examples of the MC-QTAIM analysis of polyatomic species are considered in details assuming hydrogen nuclei as quantum waves. Our model systems include some basic hydrocarbons namely methane, ethylene, acetylene and benzene as well as their congeners containing deuterons and tritons instead of protons. Within the context of the MC-QTAIM both the morphology and properties of atoms in molecules depend on the masses of all constituent quantum particles of the system [5]. Accordingly, the replacement of all protons with their heavier isotopes are discernible within the AIM analysis; this was first demonstrated in *LiX* (*X = H, D, T*) species [5]. On the other hand, the mass-dependent Gamma density is used as the basic one-particle density for the topological analysis of the MC-QTAIM [6,8] instead of the one-electron density [2-4], thus the topological analysis of each isotopically substituted species is distinct. Although the number of studied species in this contribution is not large, some patterns emerge from this contribution that based on concomitant theoretical reasoning disclose certain general traits of the MC-QTAIM analysis. One may hope that the present study and disclosed traits will pave the way for the MC-QTAIM of more complex polyatomic species.

## 2. Computational details

At first step, ab initio nBO calculations were done to derive nBO wavefunction for the MC-QTAIM analysis. The used ab initio procedure is the nuclear-electronic orbital method at Hartree-Fock level (NEO-HF) that was developed by Hammes-Schiffer and coworkers [15] where the resulting wavefunctions belong to the family of the WF1 wavefunctions as detailed previously [6]. The NEO-HF is an extended version of the orthodox Hartree-Fock procedure that from outset treats both electrons and selected nuclei as quantum waves equally. The wavefunction of the NEO-HF is a product of Slater determinants each describing one type of quantum particles (assuming all to be fermions), thus single particle functions called nuclear and electronic orbitals are attributed to selected nuclei and electrons, respectively. Therefore, the basic equations of the NEO-HF are multi-component equations and in the relevant self-consistent field (SCF) procedure both electronic and nuclear orbitals are optimized simultaneously [15]. The computer program used for the NEO-HF calculations is the NEO package developed by Hammes-Schiffer and coworkers as implemented in the GAMESS suite of programs [15,16]. In order to fit our requirements, the original NEO package was modified and new features were added as detailed below. At the next stage the produced nBO wavefunctions were used for the MC-QTAIM analysis, encompassing both the topological analysis and the basin integrations, using procedures described fully in previous publications [5,9,12]. Finally, the AIMAll package was used to visualize the Molecular Graphs (MGs) [17].

In the course of the NEO-HF calculations, all electrons and protons($H$)/deuterons($D$)/tritons($T$) were treated as quantum waves while carbon nuclei were clamped as point charges.

Throughout calculations, the used masses for the hydrogen isotopes are as follows: $H = 1836 m_e, D = 3670 m_e, T = 5496 m_e$ while to simulate the clamped nucleus limit a hypothetical heavy congener of proton with an extra-large mass, $M = 10^{20} m_e$, was used. The employed basis set for the NEO-HF calculations has both electronic and nuclear parts. The nuclear basis set is composed of single s-type Gaussian functions for each quantum nucleus that their positions and exponents were both optimized during the SCF calculations. The non-linear optimization of the exponents of basis functions is a new feature that has been added to the original NEO package. For electronic part of the basis set, the standard 6-311++G(d,p) [18] basis set was placed on the clamped carbon nuclei as well as on the center of nuclear s-type Gaussian functions used to describe quantum hydrogen nuclei. The linear coefficients of the electronic basis functions for the clamped nuclei were optimized during the SCF procedure while in the case of quantum nuclei the electronic basis set was first "de-contracted" and then both the exponents and linear coefficients of the Gaussian functions were optimized simultaneously. This procedure guarantees that the used [4s,1p] electronic basis set for hydrogen atoms, which had been designed originally just for clamped protons [18], is not only flexible enough to describe the electronic distribution around the quantum nuclei properly but also not biased. Initially, a usual HF/6-311++G(d,p) geometry optimization was done on *CH₄, C₂H₄, C₂H₂, C₆H₆* species assuming all nuclei as clamped particles. At next step, the optimized coordinates of the clamped protons were used as the initial guess for the initial position of the joint centers of the nuclear and electronic basis functions describing the quantum nuclei and surrounding electrons. During the NEO-HF calculations the joint centers as well as the positions of the clamped nuclei were all fully optimized. Accordingly, one may claim that all variables of nuclear and electronic basis functions of the quantum nuclei were optimized during the ab initio NEO-HF calculations. The usual geometrical symmetries introduced within the context of the BO paradigm namely, T$_d$ for *CX$_4$*, D$_{\infty h}$ for *C$_2$X$_2$*, D$_{2h}$ for *C$_2$X$_4$*, D$_{6h}$ for *C$_6$X$_6$ (X= H, D, T, M)* remained intact during the optimization procedure; the clamped carbon nuclei and the joint centers are now used to introduce geometrical symmetries. The spin state of electrons was assumed to be singlet throughout calculations while for nuclei all individual spins directions were assumed to be parallel yielding a total high-spin multiplet state. The fact that deuterium nucleus is a boson does not makes a difference since, as will be considered in detail in subsequent sections, the nuclear single particle functions do not overlap effectively leaving the nuclei practically "distinguishable". A new routine was added to the original NEO package enabling automatic production of the "extended" wfn files from the final fully optimized nBO wavefunction containing both electronic and nuclear orbitals. These extended wfn protocols, which their detailed structure will be disclosed in a future publication, were then used as inputs for the MC-QTAIM analysis. To check the numerical accuracy of the basin integration algorithm, the net flux integral (*vide infra*), $\tilde{L}(\Omega) = (-1/4)\int_\Omega d\vec{q}\ \nabla^2 \Gamma^{(2)}(\vec{q})$, was computed during integration of each atomic basin and in all cases it was demonstrated that $L(\Omega) < 10^{-4}$ (in atomic units). To compute basin energies, since the virial theorem is not completely satisfied (see Table 1 for the computed virial ratios), an extra ad hoc virial scaling was done as advocated previously [4,12]. All numerical results and equations are given in atomic units, throughout the paper.

## 3. Results and Discussion

### 3.1. The ab initio NEO-HF calculations

Table 1 compresses the main results of the ab initio calculations on the sixteen considered species categorized in four classes namely, $CX_4$, $C_2X_2$, $C_2X_4$, $C_6X_6$. Because of the mentioned geometrical symmetries, only a handful of geometrical parameters suffices to describe the "pseudo"-geometries of these species. The joint centers are used to introduce $C$-$X$ "mean" inter-nuclear distances concerning the fact that the distribution of each quantum nuclei is just a single s-type Gaussian function. In each class, the mean $C$-$X$ distances decrease with increasing the mass of the quantum nucleus and in the clamped nucleus limit, where one is faced practically with a Dirac delta function instead of a Gaussian function describing the hydrogen nuclei, the shortest distances are observed. This contraction of the mean inter-nuclear distances, which is a well known phenomenon [19-23], originates from the fact that heavier isotopes have smaller zero-point energies thus are more confined to the bottom of the asymmetric vibrational potential energy surface with smaller vibrational amplitudes. This interpretation is also confirmed by inspecting the optimized exponents of the nuclear Gaussian functions in each class, which are clearly larger for heavier quantum nuclei. Accordingly, a larger Gaussian exponent witnesses a more contracted nuclear distribution and smaller vibrational amplitude. The variations of the $C$-$C$ inter-nuclear distances in each class as well as the mean $H$-$C$-$H$ angle in $C_2X_4$ class are much less affected by the isotopic substitution though slight contractions upon increasing the mass of the isotopes are discernable. The total energies in each class are also clearly correlated with the mass of hydrogen isotopes and are more negative for the heavier isotopes; the most negative energies correspond to the clamped nucleus limit. This is in line with previous studies [5,6,9,14] and is rationalized based on the virial theorem that is almost satisfied inspecting the computed virial ratios given in Table 1; according to this theorem, at the equilibrium state the total energy is just the minus sum of kinetic energies of all component bodies of a system [7]. Accordingly, as is also evident from Table 2, with increasing the mass of quantum nuclei the nuclear kinetic energy decreases, being zero for the clamped nucleus. Simultaneously, because of concomitant increased localization of electronic distribution around heavier nuclei, the electronic kinetic energy increases which dominates the overall kinetic energy variations, Table 2. Roughly speaking, at the clamped nucleus limit electrons are forced to circulate faster around the point charge nuclei, being in tighter orbits than their finite mass congeners, thus having the largest electronic kinetic energy (most negative total energy) in each class.

It is also illustrative to study the trends observed in the ab initio derived data of the species with the same isotopic composition comparatively. The computed *C-H* inter-nuclear distances at the clamped nucleus limit are ordered as follows: $R_{C-H}(CM_4) > R_{C-H}(C_2M_4) \geq R_{C-H}(C_6M_6) > R_{C-H}(C_2M_2)$; this well-known trend is rationalized based on the nature of hybridized atomic orbitals on carbon atoms [24]. In contrast to the large deviations from the computed inter-nuclear distances at the clamped nucleus limit, exactly the same ordering is reproduced if one compares the mean inter-nuclear distances of species with the same isotopic composition. Interestingly, the differences in the mean *C-X* inter-nuclear distances of two species with different isotopes in the same class, e.g. *X = H, D* or *X = D, T*, is almost the same in all the classes and independent from the compared species: $\langle R_{C-H} \rangle - \langle R_{C-D} \rangle \approx 0.014$, $\langle R_{C-D} \rangle - \langle R_{C-T} \rangle \approx 0.006$, $\langle R_{C-T} \rangle - \langle R_{C-M} \rangle \approx 0.027$. The observed contraction of the mean *C-X* inter-nuclear distances with increasing the mass of hydrogen nuclei has been also well documented experimentally long ago for hydrocarbons, $\langle R_{C-H} \rangle - \langle R_{C-D} \rangle \approx 0.01$ [20-22]. The exponents of nuclear Gaussian functions ($\alpha$) describing quantum nuclei also follow a general regular pattern:

$\alpha(C_6X_6) \geq \alpha(C_2X_4) > \alpha(CX_4) > \alpha(C_2X_2)$. However, these exponents are less sensitive to *chemical environment* and reveal the nature of the used isotope unambiguously, $\alpha(H) \approx 23.4 \pm 0.5$, $\alpha(D) \approx 34.7 \pm 0.7$, $\alpha(T) \approx 43.5 \pm 0.9$, as also demonstrated in a recent study [14].

### 3.2. The Topological analysis

The basic one-particle density used for the topological analysis unraveling the AIM structure within context of the MC-QTAIM is as follows [8]:

$$\Gamma^{(P)}(\vec{q}) = \rho_1(\vec{q}) + \sum_{n=2}^{P} (m_1/m_n)\, \rho_n(\vec{q}) \tag{1}$$

In this equation $\Gamma^{(P)}(\vec{q})$ stands for the Gamma density while $\rho_n(\vec{q}) = N_n \int d\tau'_n\, \Psi\Psi^*$, $N_n, m_n$ are the one-particle density, the number of particles and the mass of the *n*-th subset of quantum particles of system under study, respectively ($\Psi$ is an ab initio nBO multi-component wavefunction while $d\tau'_n$ implies summing over spin variables of all quantum particles and integrating over spatial coordinates of all quantum particles except the coordinate of one arbitrary particle belonging to *n*-th subset denoted by $\vec{q}$). Also, $\rho_1(\vec{q})$ is the one-particle density of the subset of the lightest quantum particles of the system under study, usually electrons, with the mass $m_1$. As is evident from equation (1), the Gamma density is the mass-scaled combined density composed of the one-particle densities of all involved quantum particles. The total number of subsets, denoted as *P*, is called the cardinal number [8]. In present analysis since only electrons and one type of hydrogen isotopes are treated as quantum waves in each species then *P* = 2. Accordingly, the proper Gamma density written in atomic units is as follows:

$$\Gamma^{(2)}(\vec{q}) = \rho_e(\vec{q}) + (1/m_X)\rho_X(\vec{q}) \qquad (2)$$

In this equation the subscripts *e* and *X* stand for electrons and one of the real or the hypothetical hydrogen isotopes (*X = H, D, T, M*), respectively. Generally, the one-particle density of electrons is distinct from the orthodox one-electron density [5,6] though for *X = M*, simulating the clamped nucleus limit, there is practically no difference between these two since the second term in equation (2) vanishes and $\rho_e(\vec{q})$ is calculated assuming a Dirac delta like distribution for the quantum nuclei. Accordingly, the topological analysis of $\Gamma^{(2)}(\vec{q})$ using *X = M* for each species is indistinguishable from the topological analysis of the one-electron density of the same species performed within context of the orthodox QTAIM. The topological analysis of the Gamma density is done considering the gradient vector field of the Gamma, $\vec{\nabla}\Gamma^{(P)}(\vec{q})$, and seeking for the critical points (CPs) and the zero-flux surfaces, usually termed inter-atomic surfaces [2-4], which satisfy the local zero-flux equation, $\vec{\nabla}\Gamma^{(P)}(\vec{q}) \cdot \vec{n} = 0$ ($\vec{n}$ stands for unit vector normal to the surface) [5,6]. In the present case, *P = 2*, the gradient vector field and the local zero-flux equation are as follows:

$$\vec{\nabla}\Gamma^{(2)}(\vec{q}) = \vec{\nabla}\rho_e(\vec{q}) + (1/m_X)\vec{\nabla}\rho_X(\vec{q})$$

$$\vec{\nabla}\Gamma^{(2)}(\vec{q}) \cdot \vec{n} = \vec{\nabla}\rho_e(\vec{q}) \cdot \vec{n} + (1/m_X)\vec{\nabla}\rho_X(\vec{q}) \cdot \vec{n} = 0 \qquad (3)$$

The CPs are those points of molecular space where: $\vec{\nabla}\Gamma^{(P)}(\vec{q}_{CP}) = 0$, and are categorized using the orthodox terminology disclosed fully in Bader's monograph [2]. However, based on a recent proposal [25], a (3, -1) CP is called a "line" CP, abbreviated as LCP, instead of usual terminology calling it "bond" CP, BCP; this change of terminology paves the way for a consistent interpretation of the MC-QTAIM analysis as detailed elsewhere [25]. Accordingly, the gradient path linking a (3, -3) CP to a LCP is called a line path (LP) instead of the usually termed bond path [25]. Each combined property density, denoted as $\tilde{M}(\vec{q})$, may be computed at the LCPs; these are usually called topological indices [2-4]. A combined property density is the combination of property densities originating from all quantum particles:

$$\tilde{M}(\vec{q}) = \sum_{n=1}^{P} M_n(\vec{q}) \text{ where } M_n(\vec{q}) = N_n \int d\tau'_n \Psi^* \hat{m}_q \Psi$$ ($\hat{m}_q$ is the one-particle hermitian operator describing the property $M$); since in this study the wavefunctions are real functions, the property density are also real scalars/tensors. In the present case, $P = 2$, the topological indices are as follows:

$$\tilde{M}(\vec{q}_{LCP}) = M_e(\vec{q}_{LCP}) + M_X(\vec{q}_{LCP}) \tag{4}$$

Examples of topological indices considered in this study are the Laplacian of the Gamma, $\nabla^2 \Gamma^{(2)}(\vec{q})$, as well as the eigenvalues of its Hessian matrix, the combined Hamiltonian kinetic energy density, $\tilde{K}(\vec{q})$, the combined Lagrangian kinetic energy density, $\tilde{G}(\vec{q})$, and the combined virial density, $\tilde{V}(\vec{q})$, that all have been disclosed fully in previous contributions [7,8] and are not reiterated here.

Figure 1 and Table 3 list some results of the topological analysis. The MGs depicted in Figure 1 are quite similar to those previously emerged within the context of the orthodox QTAIM [2,3] apart from the fact that no clamped hydrogen nuclei are near the (3, -3) CPs corresponding to the hydrogen atomic basins (*vide infra*). For each class of species, just a single MG emerges; isotope substitution does not affect the qualitative patterns of these graphs. However, the quantitative analysis presented in Table 3 clearly demonstrates that both the Gamma density and the topological indices at the *C-X* associated LCPs are distinct in each species. On the other hand, the topological indices of the *C-C* associated LCPs in each class are much less affected by the isotope substitution; this is in line with the aforementioned relative insensitivity of the *C-C* mean inter-nuclear distances to the isotope substitution. A detailed dissection of both the Gamma and property densities into electronic and nuclear contributions, equations (2) and (4), demonstrates that just electrons are contributing to the LCP properties, $\Gamma^{(2)}(\vec{q}_{LCP}) = \rho_e(\vec{q}_{LCP})$ and $\tilde{M}(\vec{q}_{CP}) = M_e(\vec{q}_{CP})$. This observation hints that in all the considered species, hydrogen isotopes' distributions are "confined" to the interior of hydrogen basins, not "leaking" on or beyond the inter-atomic surfaces (*vide infra*). The fact that most sensitive topological "probes" for isotope substitution, and concomitant induced perturbation in electron's distribution, is the topological indices of the *C-X* associated LCPs conforms to the idea of the *nearsightedness of electronic matter* advocated by Bader [26,27]. Although the very nature and basic principles behind the nearsightedness are not yet quite clear [28-31], the present results indicate that it is also operative in the nBO domain.

Detailed inspection of Table 3 reveals certain patterns at the *C-X* associated LCPs that may be used to distinguish the isotopic constitution of the *C-X* bond. In the rest of this paper, in discussing these patterns, it is implicitly assumed that all regularities and trends are described from species containing proton to the hypothetical highly massive isotope thus the phrase "due to the increase of the mass" is eliminated from corresponding statements. In all classes, the length of the LP connecting the LCP to (3, -3) CP at carbon nucleus decreases whereas the reverse trend is observed for the length of the LP connecting the LCP to the (3, -3) CP resides in the hydrogen basin. These are the first evidence that in each class the hydrogen basins "expand" whereas carbon basins "shrink". Overall, the sum of the lengths of these two LPs deceases in each class; this is a sign of shrinkage of the total molecular volume in each class. The importance of these trends will be more evident when explicitly considering the atomic volumes in the next subsection. The amount of various considered topological indices generally increase in each class; these include the Gamma density which is practically the one-particle density of electrons at the LCPs (*vide supra*), the absolute amount of the Laplacian of the Gamma density, both kinetic energy densities, as well as the absolute amount of the total virial density. Also, the computed topological indices for species with *X=M* are indistinguishable from those computed within the Born-Oppenheimer paradigm employing the electronic wavefunctions and the topological analysis of the orthodox QTAIM. All in all, the "fingerprint" of hydrogen isotopes is clearly seen on the topological indices.

**3.3. Basin integration and the properties of atoms in molecules**

Each atomic basin, a topological atom, is defined by the inter-atomic surfaces originating from the local zero-flux equation, equation (3), satisfying the net zero-flux equation [8]:

$$\int_\Omega d\vec{q}\ \nabla^2 \Gamma^{(n)}(\vec{q}) = 0 \tag{5}$$

Based on a recent proposed terminology [25], atomic basins sharing an inter-atomic surface are called "neighbors". To each atomic basin, regional properties are attributed integrating the property densities [8]:

$$\tilde{M}(\Omega) = \sum_{n=1}^{P} \int_\Omega d\vec{q}\ M_n(\vec{q}) \tag{6}$$

In present study the electrons and quantum hydrogen nuclei are contributing to the regional properties:

$$\tilde{M}(\Omega) = M_e(\Omega) + M_X(\Omega) \tag{7}$$

The explicit form of each regional property used in this study has been disclosed in full detail previously and are not reiterated here [6-8]. Also, the extended theory of the localization/delocalization of electrons, which is applicable in nBO domain and articulated recently [9], is also used in this study.

The topological analysis reveals the underlying AIM structure of the considered species. With exception of the acetylene class, for each species every atomic basin contains a clamped or a quantum nucleus and the number of basins is equal to the number of nuclei, consistent with the results of the orthodox QTAIM [2-4]. In the acetylene class, also in line with previously reported orthodox QTAIM analysis [2], a (3, -3) CP appears in the middle of the two carbon nuclei and shapes a so-called "pseudo-atom" (PA) [2]. This atomic basin encompasses neither a clamped nor a quantum nucleus. Table 4 offers the population of quantum particles as well as the energy of each atomic basin where, the sum of basin energies for each species agrees well with that computed independently from the ab initio NEO-HF method. Since energy is a sensitive probe of the quality of basin integration method [4], this agreement, independent from the previously mentioned small values of the net flux integral, $\tilde{L}(\Omega)$, further strengthens the high precision of the employed numerical integration procedure. The computed population of each quantum nucleus clearly demonstrates that in all considered species each quantum nucleus is confined just to its own atomic basin without any "leakage" to the neighboring basins. This observation is in line with a recent study [14] and one may claim that the vibrational amplitudes of the quantum nuclei are not large enough to pass the boundary of their own atomic basin. The consequences of this "confinement" will be considered in detail in subsequent subsection. On the other hand, since no clamped nuclei are within the hydrogen basins, the "identity" of these basins is revealed by the population of quantum nucleus confined to that basin. The computed electronic populations of atomic basins also unravel regular patterns. Within each class, the electronic population of the hydrogen basins increases whereas that of the carbon basins decreases. Evidently, consistent with previous proposal [5,14], the electronegativity of hydrogen isotopes is correlated "directly" with their mass and the hypothetical superheavy isotope, $X=M$, has the largest electron withdrawing capacity. A comparison of species with the same isotope constitution in various classes is also instructive since it is well-known that in hydrocarbons the electronegativity of carbon atom depends on its atomic hybridization [24]; carbon atom with $sp$ hybrid is more electronegative than the carbon with $sp^2$ hybrid while the latter itself is more electronegative than $sp^3$ hybrid. The observed trends of electronic populations indeed conform well to this expectation since hydrogen basins in acetylene class have the smallest electronic populations whereas those of methane class have the largest electronic populations and those belong to the ethylene and benzene classes are in the middle range and quite similar to each other. In each class, there is a marked direct correlation between the magnitude of the basin energies and their electronic population, including PAs, which has been observed in our recent study [14]. Table 5 offers the volumes of all considered basins that have been introduced recently as a regional property [9]:

$$V(\Omega) = \int_\Omega k \, d\vec{q}, \quad k = \begin{cases} 1, & \text{if } \Gamma^{(2)}(\vec{q}) \geq 0.001 \\ 0, & \text{if } \Gamma^{(2)}(\vec{q}) \prec 0.001 \end{cases} \quad (8)$$

Practically, since the quantum nuclei are totally confined within their own atomic basins in the considered species, the electron one-particle density, $\rho_e(\vec{q})$, determines the "outer" boundary of atomic basins; the *one-electron density* has the same role within context of the orthodox QTAIM [32]. In agreement with the orthodox view [19-23,33-35], the molecular volume decreases in each class however, in contrast to the orthodox view, the volumes of hydrogen basins *expand* in each class and the reason behind shrinkage of molecular volumes is the larger *shrinkage* of the carbon basins. Although, it has recently been speculated that deuterium containing hydrogen atoms may have a larger effective size than proton containing hydrogen atoms in certain cases [36], according to the best of authors' knowledge, this is the first report directly challenging the orthodox view on the relative size of proton and deuterium containing hydrogen atoms. Interestingly, from the previous MC-QTAIM analysis on hydrogen molecule and its isotopomers ($H_2$, $D_2$, $T_2$) [9], in line with orthodox view, the emerging volume ordering was as follows: $V_H(\Omega) > V_D(\Omega) > V_T(\Omega)$. This demonstrates that the reverse observed ordering in the present study: $V_H(\Omega) < V_D(\Omega) < V_T(\Omega)$, emerges probably from the increase in electronic population and larger electron withdrawing capacity of the heavier isotopes. These quite interesting observations need further verifications in future studies. The electronic localization (EL) and delocalization (ED) indices of neighboring basins in the considered species are gathered in Table 6. The results obtained for $X=M$ in all classes are numerically quite similar to those reported previously within the context of the orthodox QTAIM's localization/delocalization theory [37]. In each class both ELs and EDs are affected by isotope substitution though the EDs are less sensitive to these substitutions. It is evident from the table that the ELs of the carbon basins in each class decrease whereas those of the hydrogen basins increase; this is better seen in the percent localization variations demonstrating the larger capacity of heavier isotopes to localize electrons in their basins. Table 7 offers specially various carbon-carbon EDs in the benzene class since previously they have been used as indicators of the π-aromaticity of benzene and larger 6-membered polycyclic hydrocarbons [38,39]. In agreement with previous reports [38,39], the EDs between the para carbons are larger than those between the meta carbons in all the four considered species revealing the fact that replacing quantum instead of the clamped hydrogen nuclei does not change the expected pattern of EDs in benzene molecule.

One may conclude from this subsection that the fingerprint of quantum nuclei, though usually as a *fine* structure, is seen in all the computationally considered regional properties of topological atoms. In some cases like basin energies, quantum nuclei contribute "directly", equation (7). However, even when they are not contributing directly, e.g. topological indices at LCPs or EL/ED indices, their "indirect" influence is observable through shaping the electronic distribution. This is comprehensible since in the SCF procedure of the NEO-HF method electrons are "sensing" different environments, i.e. mean effective potential, when encountering quantum nuclei with different masses. The resulting mass-dependent "fingerprint" remains on the electronic part, e.g. Slater determinant and one-electron functions, of the nBO wavefunction. Although the electronic parts of wavefunctions are heavily processed in the MC-QTAIM analysis, evidently, they still carry the fingerprint of quantum nuclei in all aspects of the MC-QTAIM analysis.

**3.4. The consequences of confinement of quantum nucleus in a single atomic basin**

The confinement of the each quantum nucleus in its own basin may be used to simplify the MC-QTAIM formalism as is discussed briefly in this subsection. Indeed in a recent study Hammes-Schiffer and coworkers have computationally considered cases that all/selected quantum protons of a molecule are localized in certain non-overlapping regions of molecular space, each described by a nuclear orbital [40]. In such cases, the quantum exchange term in the NEO-HF equations are negligible so they have proposed that one may use the Hartree product of nuclear orbitals, instead of the Slater determinant, from outset for the nuclear part of the wavefunction. In this reformulation of the mean field equations, since the nuclear orbitals have no effective overlap, practically, quantum nuclei are *distinguishable* particles. The computational MC-QTAIM analysis in present study also conforms well to this picture and one may claim that each quantum nucleus may be labeled as a distinguishable particle. Accordingly, instead of two types of quantum particles in the considered systems, there are *N+1* quantum particles where *N* is the number of the "localized" distinguishable quantum nuclei. The equations for the Gamma and property densities are modified as follows:

$$\Gamma^{(N+1)}(\vec{q}) = \rho_e(\vec{q}) + \sum_{p=1}^{N}(1/m_{Xp})\rho_{Xp}(\vec{q})$$

$$\tilde{M}(\vec{q}) = M_e(\vec{q}) + \sum_{p=1}^{N} M_{Xp}(\vec{q}) \qquad (9)$$

In these equations the index *p* labels distinguishable quantum nuclei (it must be distinguished from the capital p, *P*, employed previously for the cardinal number). Since each quantum nucleus resides in the interior of a single basin, after integration of equations (9) the regional properties are as follows:

$$\tilde{M}(\Omega_p) = M_e(\Omega_p) + M_{Xp}(\Omega_p) \qquad (10)$$

This equation asserts that for the basin $\Omega_p$ (now the subscript *p* also enumerates the atomic basin containing the *p-th* quantum nucleus) just electrons and one of quantum nuclei, denoted by *Xp*, contribute to the basin properties and the remaining *N-1* quantum nuclei are not directly contributing in basin properties, $M_{Xj \neq p}(\Omega_p) = 0$. This is best illustrated in the case of nuclear populations, $N_{Xp}(\Omega_p)$, as well as the localization, $\lambda_{Xp}(\Omega_p)$, and the delocalization, $\delta_{Xp}(\Omega_p, \Omega_j)$, indices [9]:

$$N_{Xp}(\Omega_p) = 1, \qquad N_{Xp}(\Omega_{j \neq p}) = 0, \qquad N_{Xj \neq p}(\Omega_p) = 0,$$

$$\lambda_{Xp}(\Omega_p) = 1, \qquad \delta_{Xp}(\Omega_p, \Omega_{j \neq p}) = 0 \qquad (11)$$

These equations "quantify" the localization of the quantum nuclei within the basin $\Omega_p$. The localization may have interesting consequences as is discussed first in the case of the regional virial theorem.

The local form of the virial theorem for a region containing just two types of quantum particles, electrons and a quantum nucleus, is written as follows [7]:

$$2\tilde{T}(\vec{q}) + \tilde{V}^b(\vec{q}) = -\tilde{V}^s(\vec{q}) - (1/4)\nabla^2 \Gamma^{(2)}(\vec{q}) \qquad (12)$$

In this equation $\tilde{T}(\vec{q}), \tilde{V}^b(\vec{q})$ and $\tilde{V}^s(\vec{q})$ are the Hamiltonian kinetic, basin virial and surface virial densities, respectively (for a detailed discussion on each term see [7]). Integrating this equation for the atomic basin containing solely both types of quantum particles one arrives at:

$$2\tilde{T}(\Omega_p) + \tilde{V}^b(\Omega_p) = -\oint_{\partial\Omega p} dS(\Omega_p, \vec{q}) \; \vec{q} \cdot (\tilde{\sigma}(\vec{q}) \bullet \vec{n}) - (1/4) \int_{\Omega p} d\vec{q} \; \nabla^2 \Gamma^{(2)}(\vec{q}) \qquad (13)$$

where $\tilde{\sigma}(\vec{q}) = \tilde{\sigma}_e(\vec{q}) + \sum_{p=1}^{N} \tilde{\sigma}_{Xp}(\vec{q})$ stands for the joint stress tensor density and the symbol $\bullet$ has been used to stress on the dyadic/tensor nature of the product (for a detailed discussion on each term see [7]). Since the net (or alternatively local) zero-flux equation of the Gamma density, equation (5), is satisfied, the second term in the right hand-side of equation (13) vanishes and one arrives at the regional virial theorem:

$$2\tilde{T}(\Omega_p) = -\tilde{V}^T(\Omega_p) \; , \quad \tilde{V}^T(\Omega_p) = \tilde{V}^b(\Omega_p) + \tilde{V}^s(\Omega_p) \qquad (14)$$

As has also previously been stressed [11], equation (12) is a combination of two *separate* local virial theorems for electrons and the quantum nucleus:

$$2T_e(\vec{q}) + V_e^b(\vec{q}) = -V_e^s(\vec{q}) - (1/4)\nabla^2 \rho_e(\vec{q})$$

$$2T_{Xp}(\vec{q}) + V_{Xp}^b(\vec{q}) = -V_{Xp}^s(\vec{q}) - (1/4m_{Xp})\nabla^2 \rho_{Xp}(\vec{q}) \qquad (15)$$

The logic behind the *principle of density combination* as the cornerstone of the MC-QTAIM formalism is the fact that the regional integration of these two equations yield two separate zero-flux equations that are not satisfied simultaneously in general [7]. However, because of the assumed confinement of the nuclear orbital within the atomic basin, the net zero-flux integral for the quantum nuclei is also assumed to be practically null within its own basin, $\int_{\Omega_p} d\vec{q}\; \nabla^2 \rho_{Xp}(\vec{q}) = 0$. On the other hand, because the zero-flux equation of the Gamma density, equation (5), is satisfied, one automatically derives: $\int_{\Omega_p} d\vec{q}\; \nabla^2 \rho_e(\vec{q}) = 0$. Thus, in the special case considered here the integration of equations (15) also yield two separate regional virial theorems:

$$2T_e(\Omega_p) + \tilde{V}_e^b(\Omega_p) = -\oint_{\partial\Omega_p} dS(\Omega_p, \vec{q})\; \vec{q} \cdot (\vec{\tilde\sigma}_e(\vec{q}) \bullet \vec{n})$$

$$2T_{Xp}(\Omega_p) + \tilde{V}_{Xp}^b(\Omega_p) = -\oint_{\partial\Omega_p} dS(\Omega_p, \vec{q})\; \vec{q} \cdot (\vec{\tilde\sigma}_{Xp}(\vec{q}) \bullet \vec{n}) = 0 \qquad (16)$$

The right hand-side of the second equation is null since it is assumed that because of the confinement, the nuclear contribution to the stress tensor density is also effectively null at the atomic boundaries or in other words, the nuclear contribution of the surface virial vanishes [7]. The "disentanglement" of the nuclear and electronic regional virial theorems means that the nuclear and electronic contribution of basin energies may be computed separately and then added yielding the total basin energy as also stressed previously [7].

The role of the confinement may be now generalized considering the local hypervirial theorem, as the basic equation of the MC-QTAIM [8]:

$$\tilde{M}(\vec{q}) = \vec{\nabla} \bullet \tilde{J}_G(\vec{q}) \qquad (17)$$

In this equation, the property $M$ is generated by the one-particle generator $g$, which is a one-particle hermitian operator, through the equation: $(i)[\hat{h}_q, \hat{g}_q] = \hat{m}_q$, where $\hat{h}_q$ is part of the Hamiltonian containing spatial variables of the particle denoted by the coordinate $\vec{q}$ (for a detailed discussion see [8]). Also, the right hand-side of equation (17) contains the total current property density which is the combination of the current property densities of electrons and quantum nuclei, $\tilde{J}_G(\vec{q}) = \vec{J}_G^e(\vec{q}) + \sum_{p=1}^{N} \vec{J}_G^{Xp}(\vec{q})$, where $\vec{J}_G^e(\vec{q}) = (N_e/2i) \int d\tau'_e \{\Psi^* \vec{\nabla}_q (\hat{g}_q \Psi) - (\hat{g}_q \Psi)(\vec{\nabla}_q \Psi^*)\}$ is the electronic contribution and $\vec{J}_G^{Xp}(\vec{q}) = (1/2im_{Xp}) \int d\tau'_{Xp} \{\Psi^* \vec{\nabla}_q (\hat{g}_q \Psi) - (\hat{g}_q \Psi)(\vec{\nabla}_q \Psi^*)\}$ is the $p$-th nuclear contribution. Equation (17) is the result of the combination of $N+1$ separate local hypervirial theorems for electrons and quantum nuclei:

$$M_e(\vec{q}) = \vec{\nabla} \bullet \vec{J}_G^e(\vec{q})$$

$$M_{Xp}(\vec{q}) = \vec{\nabla} \bullet \vec{J}_G^{Xp}(\vec{q}), \qquad p = 1,...,N \qquad (18)$$

However, assuming that nuclear current property densities are effectively encompassed in their own basins, the right hand-side of the regional nuclear hypervirial theorem is practically null (in principle, it is feasible to "construct" a group of generators that corresponding current densities are not null even if the nuclear orbital is totally encompassed in an atomic basin however, the usual generators used to derive familiar regional theorems do not belong to this group [7]). Then, the regional hypervirial theorems for *p-th* atomic basin containing *p-th* localized quantum nucleus are as follows:

$$\text{Re}\left\{\int_{\Omega p} d\vec{q}\ M_e(\vec{q})\right\} = \text{Re}\left\{\oint_{\partial\Omega p} dS(\Omega p, \vec{q})\ \vec{J}_G^e(\vec{q})\cdot\vec{n}\right\}$$

$$\text{Re}\left\{\int_{\Omega p} d\vec{q}\ M_{Xp}(\vec{q})\right\} = 0 \qquad (19)$$

Assuming $\hat{g} = \hat{q}\cdot\hat{p}$ in these equations, where $\hat{p}$ stands for the linear momentum operator, equations (16) are retrieved from equations (19) while as another example by applying $\hat{g} = \hat{p}$ it emerges that the Ehrenfest forces operative on basin boundaries just arise from electrons and the nuclear contribution is null: $\tilde{F}(\Omega_i) = F_e(\Omega_i)$. Equations (19) are the basis of future developments of the MC-QTAIM for molecular systems containing localized quantum nuclei. However, for systems containing light particles that penetrate into neighboring basins, e.g. positrons or even muons in some cases [11-14], or systems with intra-molecular hydrogen tunneling, e.g. malonaldehyde [41], the general local hypervirial theorem, namely equation (17), must be used.

## 4. Prospects

In all previous computational studies only diatomic species were considered within context of the MC/TC-QTAIM. However, present study demonstrates that the MC-QTAIM is also capable of dealing with polyatomic species. The relevant reported technical developments, both computational and theoretical, open the door for the AIM analysis of a large number of polyatomic species. Particularly, ongoing progress in nBO ab initio procedures as well as the development of user-friendly computer codes like the NEO and the LOWDIN packages (the latter seems to be released in near future [42]) promises that the MC-QTAIM analysis may found widespread applications after all. For systems containing massive nuclei, the MC-QTAIM analysis marginally alters the results of the orthodox QTAIM as stressed elsewhere [7]. However, for systems containing hydrogen nuclei as well as exotic particles like muons and positrons, the MC-QTAIM analysis reveals totally new features with no analogues in the orthodox analysis.

Molecular systems containing hydrogen bonds are interesting targets for the MC-QTAIM analysis since recently particular attention has been paid to the "quantum" nature of hydrogen nuclei in this type of bonding [43-47]. If proton is treated as a quantum wave instead of a clamped nucleus in ab initio procedures, then the MC-QTAIM analysis may reveal "chemical" consequences of the "fuzziness" of quantum proton [48]. Additionally, the consequences of the isotope substitution on hydrogen bond properties may be considered by replacing quantum proton with quantum deuterium and then performing the MC-QTAIM analysis. Both cases of inter- and intra-molecular proton tunneling [49-52] are also another interesting areas for the MC-QTAIM analysis particularly since tunneling influences directly the proton transfer mechanism [53-56]. Various isotopes effects considered traditionally in physical organic chemistry are also promising fields of research for future studies. However, probably the most interesting possibility is the MC-QTAIM analysis of quantum states that are the *superposition* of two or more distinct nuclear configurations; in the case malonaldehyde as a typical example, the distribution of a quantum proton instead of being localized around one of the oxygen atoms, is evenly distributed between the two oxygen atoms [41]. Such bizarre quantum superposition, usually called "Schrödinger cat" state, has been observed experimentally [57] and there are speculations on its "chemical" implications in hydrogen bonded systems [58]. All these systems promise that the MC-QTAIM analysis in the case of polyatomic systems may yield totally new results with not even remote analogues within context of the orthodox QTAIM.

## Acknowledgments

The authors are grateful to Masumeh Gharabaghi, Cina Foroutan-Nejad and Rohoullah Firouzi for their detailed reading of a previous draft of this paper and helpful suggestions.

Figure captions:

Figure 1- The MGs of the considered species. Since the isotope substitution does not change the MGs, a single MG is depicted for each series. The larger grays spheres are the clamped carbon nuclei, superimposed on the corresponding (3, -3) CPs. The smaller green, red and yellow spheres are LCPs, (3, -3) CPs and ring CP (RCP), respectively. The black lines are LPs while the red sphere in the middle of acetylene denotes a non-nuclear attractor.

Table 1- Some results of the ab initio NEO-HF calculations including mean inter-nuclear distances (C-H), inter-nuclear distances (C-C), mean bond angles (H-C-H), total energy, virial ratios (the minus of the ratio of total potential energy to the total kinetic energy), the exponents of the nuclear s-type Gaussian functions in the basis set (nuclear exponents). All results are given in atomic units.[*]

|  | C-H | energy | virial ratio | nuclear exponents |  |  |
|---|---|---|---|---|---|---|
| $CH_4$ | 2.092 | -40.0409 | 2.0003 | 23.5 |  |  |
| $CD_4$ | 2.078 | -40.0870 | 2.0003 | 34.8 |  |  |
| $CT_4$ | 2.072 | -40.1080 | 2.0003 | 43.6 |  |  |
| $CM_4$ | 2.045 | -40.2093 | 2.0003 | -- |  |  |
|  | C-H | C-C | energy | virial ratio | nuclear exponents |  |
| $C_2H_2$ | 2.038 | 2.239 | -76.7611 | 2.0004 | 22.9 |  |
| $C_2D_2$ | 2.025 | 2.238 | -76.7836 | 2.0004 | 34.0 |  |
| $C_2T_2$ | 2.019 | 2.237 | -76.7939 | 2.0004 | 42.6 |  |
| $C_2M_2$ | 1.992 | 2.236 | -76.8434 | 2.0004 | -- |  |
|  | C-H | C-C | H-C-H | energy | virial ratio | nuclear exponents |
| $C_2H_4$ | 2.079 | 2.494 | 116.74 | -77.8877 | 2.0004 | 23.6 |
| $C_2D_4$ | 2.065 | 2.493 | 116.73 | -77.9339 | 2.0004 | 35.0 |
| $C_2T_4$ | 2.059 | 2.493 | 116.73 | -77.9550 | 2.0004 | 43.8 |
| $C_2M_4$ | 2.031 | 2.492 | 116.71 | -78.0567 | 2.0004 | -- |
|  | C-H | C-C | energy | virial ratio | nuclear exponents |  |
| $C_6H_6$ | 2.076 | 2.621 | -230.5033 | 2.0005 | 23.7 |  |
| $C_6D_6$ | 2.062 | 2.621 | -230.5729 | 2.0005 | 35.1 |  |
| $C_6T_6$ | 2.056 | 2.621 | -230.6048 | 2.0005 | 43.9 |  |
| $C_6M_6$ | 2.029 | 2.620 | -230.7579 | 2.0005 | -- |  |

[*] The mean C-H distance is the distance between the clamped carbon nucleus and the center of s-type nuclear Gaussian function describing quantum hydrogen nucleus.

Table 2- The ab initio computed electronic, nuclear and total kinetic energies. All results are given in atomic units.

|     | electronic | nuclear | total |
|-----|------------|---------|-------|
| **CH$_4$** | 39.9514 | 0.0768 | 40.0282 |
| **CD$_4$** | 40.0174 | 0.0569 | 40.0743 |
| **CT$_4$** | 40.0479 | 0.0476 | 40.0955 |
| **CM$_4$** | 40.1973 | 0.0000 | 40.1973 |
|     |          |        |        |
| **C$_2$H$_2$** | 76.6924 | 0.0375 | 76.7299 |
| **C$_2$D$_2$** | 76.7246 | 0.0278 | 76.7524 |
| **C$_2$T$_2$** | 76.7395 | 0.0232 | 76.7628 |
| **C$_2$M$_2$** | 76.8125 | 0.0000 | 76.8125 |
|     |          |        |        |
| **C$_2$H$_4$** | 77.7770 | 0.0772 | 77.8542 |
| **C$_2$D$_4$** | 77.8434 | 0.0572 | 77.9006 |
| **C$_2$T$_4$** | 77.8740 | 0.0478 | 77.9218 |
| **C$_2$M$_4$** | 78.0236 | 0.0000 | 78.0236 |
|     |          |        |        |
| **C$_6$H$_6$** | 230.2766 | 0.1163 | 230.3929 |
| **C$_6$D$_6$** | 230.3768 | 0.0861 | 230.4629 |
| **C$_6$T$_6$** | 230.4229 | 0.0720 | 230.4949 |
| **C$_6$M$_6$** | 230.6486 | 0.0000 | 230.6486 |

Table 3- Some results of the topological analysis including the Gamma density and the topological indices as well as the LP lengths (the distance between (3, -3) CP and the corresponding LCP). Since there is more than a single type of LCP in some classes, to delineate the type, italic headlines are used in the first column denoting the two atoms sharing the LCP. All results are given in atomic units.

| *C-X* | LP length C-LCP | LP length X-LCP | Γ | Lap. Γ | G | K | V | Hessian eigenvalues[*] deg. | non-deg. | |
|---|---|---|---|---|---|---|---|---|---|---|
| CH$_4$ | 1.356 | 0.694 | 0.269 | -1.042 | 0.036 | 0.297 | -0.333 | -0.676 | 0.309 | |
| CD$_4$ | 1.320 | 0.725 | 0.274 | -1.053 | 0.040 | 0.303 | -0.344 | -0.688 | 0.323 | |
| CT$_4$ | 1.305 | 0.738 | 0.276 | -1.053 | 0.042 | 0.305 | -0.347 | -0.693 | 0.334 | |
| CM$_4$ | 1.256 | 0.755 | 0.286 | -1.051 | 0.049 | 0.312 | -0.360 | -0.719 | 0.386 | |
| | | | | | | | | | | |
| *C-X* | C-LCP | X-LCP | Γ | Lap. Γ | G | K | V | deg. | non-deg. | |
| C$_2$H$_2$ | 1.434 | 0.554 | 0.282 | -1.249 | 0.021 | 0.334 | -0.355 | -0.808 | 0.367 | |
| C$_2$D$_2$ | 1.395 | 0.590 | 0.288 | -1.271 | 0.024 | 0.342 | -0.365 | -0.820 | 0.369 | |
| C$_2$T$_2$ | 1.377 | 0.607 | 0.291 | -1.270 | 0.025 | 0.343 | -0.368 | -0.825 | 0.380 | |
| C$_2$M$_2$ | 1.317 | 0.641 | 0.302 | -1.208 | 0.030 | 0.332 | -0.363 | -0.852 | 0.496 | |
| | | | | | | | | | | |
| *C-C* | C-LCP[**] | C-LCP[**] | Γ | Lap. Γ | G | K | V | deg. | non-deg. | |
| C$_2$H$_2$ | 0.940 | 1.299 | 0.425 | -1.278 | 0.367 | 0.686 | -1.053 | -0.678 | 0.078 | |
| C$_2$D$_2$ | 0.939 | 1.298 | 0.425 | -1.274 | 0.368 | 0.686 | -1.054 | -0.676 | 0.078 | |
| C$_2$T$_2$ | 0.940 | 1.298 | 0.425 | -1.272 | 0.368 | 0.686 | -1.054 | -0.675 | 0.078 | |
| C$_2$M$_2$ | 0.942 | 1.294 | 0.424 | -1.268 | 0.367 | 0.684 | -1.052 | -0.671 | 0.075 | |
| | | | | | | | | | | |
| *C-X* | C-LCP | X-LCP | Γ | Lap. Γ | G | K | V | non-deg. | non-deg. | non-deg. |
| C$_2$H$_4$ | 1.371 | 0.664 | 0.277 | -1.121 | 0.030 | 0.310 | -0.340 | -0.729 | -0.720 | 0.328 |
| C$_2$D$_4$ | 1.335 | 0.695 | 0.283 | -1.133 | 0.033 | 0.317 | -0.350 | -0.742 | -0.734 | 0.343 |
| C$_2$T$_4$ | 1.320 | 0.707 | 0.285 | -1.134 | 0.035 | 0.318 | -0.353 | -0.748 | -0.740 | 0.354 |
| C$_2$M$_4$ | 1.270 | 0.726 | 0.295 | -1.127 | 0.041 | 0.323 | -0.363 | -0.776 | -0.769 | 0.418 |
| | | | | | | | | | | |
| *C-C* | C-LCP | C-LCP | Γ | Lap. Γ | G | K | V | non-deg. | non-deg. | non-deg. |
| C$_2$H$_4$ | 1.247 | 1.247 | 0.358 | -1.178 | 0.140 | 0.435 | -0.575 | -0.788 | -0.567 | 0.177 |
| C$_2$D$_4$ | 1.247 | 1.247 | 0.358 | -1.181 | 0.140 | 0.435 | -0.575 | -0.790 | -0.567 | 0.177 |
| C$_2$T$_4$ | 1.247 | 1.247 | 0.358 | -1.182 | 0.140 | 0.435 | -0.575 | -0.791 | -0.567 | 0.177 |
| C$_2$M$_4$ | 1.246 | 1.246 | 0.359 | -1.185 | 0.140 | 0.436 | -0.576 | -0.795 | -0.567 | 0.176 |
| | | | | | | | | | | |
| *C-X* | C-LCP | X-LCP | Γ | Lap. Γ | G | K | V | non-deg. | non-deg. | non-deg. |
| C$_6$H$_6$ | 1.368 | 0.665 | 0.279 | -1.132 | 0.030 | 0.313 | -0.343 | -0.740 | -0.725 | 0.332 |
| C$_6$D$_6$ | 1.333 | 0.694 | 0.285 | -1.142 | 0.033 | 0.319 | -0.352 | -0.753 | -0.738 | 0.349 |
| C$_6$T$_6$ | 1.319 | 0.707 | 0.287 | -1.142 | 0.035 | 0.320 | -0.355 | -0.758 | -0.744 | 0.361 |
| C$_6$M$_6$ | 1.271 | 0.724 | 0.297 | -1.139 | 0.040 | 0.325 | -0.365 | -0.786 | -0.773 | 0.420 |
| | | | | | | | | | | |
| *C-C* | C-LCP | C-LCP | Γ | Lap. Γ | G | K | V | non-deg. | non-deg. | non-deg. |
| C$_6$H$_6$ | 1.311 | 1.311 | 0.323 | -1.019 | 0.098 | 0.353 | -0.451 | -0.688 | -0.563 | 0.232 |
| C$_6$D$_6$ | 1.311 | 1.311 | 0.323 | -1.021 | 0.098 | 0.353 | -0.451 | -0.689 | -0.563 | 0.232 |
| C$_6$T$_6$ | 1.310 | 1.310 | 0.323 | -1.021 | 0.098 | 0.353 | -0.451 | -0.690 | -0.564 | 0.232 |
| C$_6$M$_6$ | 1.310 | 1.310 | 0.324 | -1.024 | 0.098 | 0.354 | -0.452 | -0.691 | -0.564 | 0.231 |

[*] deg. stands for degenerate while non-deg. for non-degenerate.
[**] Since there is a non-nuclear attractor in the middle of the carbon-carbon bond, there are two LCPs connecting each (3, -3) on the carbon nuclei to the non-nuclear attractor. The two reported lengths are the distances of carbon nucleus from each of the LCPs.

Table 4- Some results of the basin integrations including basin energies, populations of quantum nuclei (denoted as X) and electrons in each basin. All results are given in atomic units.

|  | basin energy | | | | X population | | electron population | | |
| --- | --- | --- | --- | --- | --- | --- | --- | --- | --- |
|  | C | X | PA | Sum* | C | X | C | X | PA |
| $CH_4$ | -37.7693 | -0.5679 |  | -40.0409 | 0.000 | 1.000 | 6.153 | 0.962 |  |
| $CD_4$ | -37.7194 | -0.5919 |  | -40.0870 | 0.000 | 1.000 | 6.022 | 0.995 |  |
| $CT_4$ | -37.6986 | -0.6023 |  | -40.1080 | 0.000 | 1.000 | 5.967 | 1.008 |  |
| $CM_4$ | -37.6379 | -0.6428 |  | -40.2093 | 0.000 | 1.000 | 5.787 | 1.053 |  |
|  |  |  |  |  |  |  |  |  |  |
| $C_2H_2$ | -37.6573 | -0.4675 | -0.5114 | -76.7610 | 0.000 | 1.000 | 5.855 | 0.725 | 0.840 |
| $C_2D_2$ | -37.6423 | -0.4935 | -0.5119 | -76.7835 | 0.000 | 1.000 | 5.814 | 0.766 | 0.841 |
| $C_2T_2$ | -37.6359 | -0.5055 | -0.5110 | -76.7938 | 0.000 | 1.000 | 5.796 | 0.785 | 0.839 |
| $C_2M_2$ | -37.6215 | -0.5497 | -0.5008 | -76.8433 | 0.000 | 1.000 | 5.744 | 0.845 | 0.823 |
|  |  |  |  |  |  |  |  |  |  |
| $C_2H_4$ | -37.8232 | -0.5603 |  | -77.8877 | 0.000 | 1.000 | 6.142 | 0.929 |  |
| $C_2D_4$ | -37.7985 | -0.5842 |  | -77.9339 | 0.000 | 1.000 | 6.076 | 0.962 |  |
| $C_2T_4$ | -37.7882 | -0.5946 |  | -77.9550 | 0.000 | 1.000 | 6.048 | 0.976 |  |
| $C_2M_4$ | -37.7580 | -0.6351 |  | -78.0566 | 0.000 | 1.000 | 5.954 | 1.023 |  |
|  |  |  |  |  |  |  |  |  |  |
| $C_6H_6$ | -37.8515 | -0.5657 |  | -230.5032 | 0.000 | 1.000 | 6.068 | 0.932 |  |
| $C_6D_6$ | -37.8396 | -0.5892 |  | -230.5728 | 0.000 | 1.000 | 6.035 | 0.965 |  |
| $C_6T_6$ | -37.8347 | -0.5994 |  | -230.6047 | 0.000 | 1.000 | 6.021 | 0.979 |  |
| $C_6M_6$ | -37.8206 | -0.6391 |  | -230.7579 | 0.000 | 1.000 | 5.976 | 1.024 |  |

*The sum is as follows: $E_t = nE_C(\Omega) + mE_X(\Omega) + pE_{PA}(\Omega)$ where $m, n, p$ stand for the number of carbon, hydrogen and pseudo-atom basins.

Table 5- The computed atomic volumes all offered in atomic units.

|  | C | X | PA | sum |
|---|---|---|---|---|
| **$CH_4$** | 91.2 | 51.1 |  | 295.7 |
| **$CD_4$** | 86.3 | 51.6 |  | 292.7 |
| **$CT_4$** | 84.3 | 51.8 |  | 291.3 |
| **$CM_4$** | 77.5 | 51.7 |  | 284.5 |
|  |  |  |  |  |
| **$C_2H_2$** | 124.9 | 38.5 | 24.0 | 350.7 |
| **$C_2D_2$** | 123.1 | 39.5 | 24.1 | 349.4 |
| **$C_2T_2$** | 122.4 | 40.0 | 24.1 | 348.9 |
| **$C_2M_2$** | 119.4 | 41.1 | 23.5 | 344.6 |
|  |  |  |  |  |
| **$C_2H_4$** | 104.9 | 49.2 |  | 406.7 |
| **$C_2D_4$** | 102.5 | 49.6 |  | 403.4 |
| **$C_2T_4$** | 101.5 | 49.8 |  | 402.4 |
| **$C_2M_4$** | 98.1 | 50.1 |  | 396.4 |
|  |  |  |  |  |
| **$C_6H_6$** | 87.5 | 48.7 |  | 817.4 |
| **$C_6D_6$** | 86.2 | 49.1 |  | 811.8 |
| **$C_6T_6$** | 85.6 | 49.3 |  | 809.3 |
| **$C_6M_6$** | 83.9 | 49.5 |  | 800.3 |

Table 6- The computed electronic localization and delocalization indices as well as the percent localization.

| | electronic localization | | | percent localization | | | electronic delocalization | | |
|---|---|---|---|---|---|---|---|---|---|
| | C | X | PA | C | X | PA | C-X | | |
| $CH_4$ | 4.19 | 0.41 | | 68.1 | 43.1 | | 0.98 | | |
| $CD_4$ | 4.06 | 0.44 | | 67.3 | 44.5 | | 0.98 | | |
| $CT_4$ | 4.00 | 0.45 | | 67.0 | 45.1 | | 0.98 | | |
| $CM_4$ | 3.82 | 0.50 | | 66.1 | 47.0 | | 0.98 | | |
| | | | | | | | C-C | C-X | C-PA |
| $C_2H_2$ | 3.97 | 0.24 | 0.14 | 67.8 | 33.3 | 16.2 | 2.14 | 0.90 | 0.69 |
| $C_2D_2$ | 3.92 | 0.27 | 0.14 | 67.2 | 35.8 | 16.2 | 2.13 | 0.92 | 0.69 |
| $C_2T_2$ | 3.90 | 0.28 | 0.14 | 67.2 | 35.9 | 16.2 | 2.13 | 0.93 | 0.69 |
| $C_2M_2$ | 3.83 | 0.32 | 0.13 | 66.7 | 38.4 | 15.8 | 2.13 | 0.96 | 0.67 |
| $C_2H_4$ | 4.16 | 0.39 | | 67.6 | 41.9 | | 1.92 | 0.96 | |
| $C_2D_4$ | 4.08 | 0.42 | | 67.2 | 43.3 | | 1.91 | 0.97 | |
| $C_2T_4$ | 4.06 | 0.43 | | 67.1 | 43.8 | | 1.90 | 0.98 | |
| $C_2M_4$ | 3.96 | 0.47 | | 66.5 | 45.7 | | 1.88 | 0.98 | |
| $C_6H_6$ | 4.01 | 0.39 | | 66.1 | 42.1 | | 1.40 | 0.96 | |
| $C_6D_6$ | 3.98 | 0.42 | | 65.9 | 43.4 | | 1.40 | 0.96 | |
| $C_6T_6$ | 3.96 | 0.43 | | 65.8 | 43.9 | | 1.39 | 0.97 | |
| $C_6M_6$ | 3.92 | 0.47 | | 65.5 | 45.7 | | 1.39 | 0.98 | |

Table 7- The computed delocalization indices between various carbon basins (C-C) of the benzene class.

|  | ortho | meta | para |
|---|---|---|---|
| $C_6H_6$ | 1.400 | 0.075 | 0.101 |
| $C_6D_6$ | 1.396 | 0.074 | 0.100 |
| $C_6T_6$ | 1.394 | 0.074 | 0.100 |
| $C_6M_6$ | 1.388 | 0.074 | 0.099 |

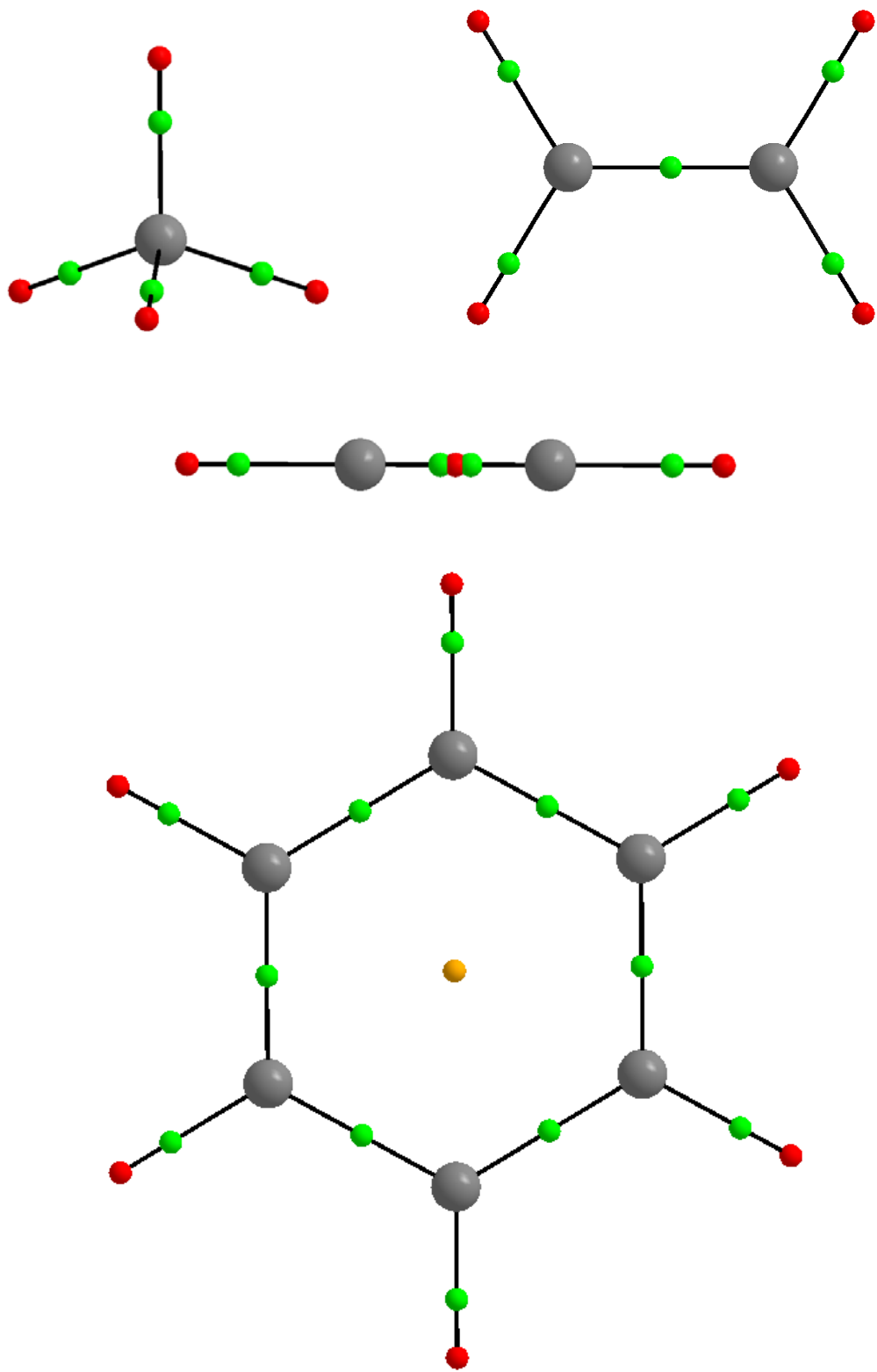

Figure-1